\documentclass[aps,prd,amsmath,amssymb,showpacs]{revtex4}

\usepackage{graphicx}
\begin{document}

\title{Realistic  search for  doubly charged bileptons 
at linear $e^{-}e^{-}$ collider energies}
\author{S. Ata\u{g}}
\email[]{atag@science.ankara.edu.tr}
\affiliation{Department of Physics, Faculty of Sciences,
Ankara University, 06100 Tandogan, Ankara, Turkey}

\author{K.O. Ozansoy }
\email[]{oozansoy@science.ankara.edu.tr}
\affiliation{Department of Physics, Faculty of Sciences,
Ankara University, 06100 Tandogan, Ankara, Turkey}

\begin{abstract}
We search for doubly charged scalar bileptons via
$ \text{e}^{-}\text{e}^{-}\to \mu^{-}\mu^{-}$ and 
$ \text{e}^{-}\text{e}^{-}\to \text{e}^{-}\text{e}^{-}$ 
processes at linear collider energies by considering 
initial and final state electromagnetic radiative 
corrections (ISR, FSR). Moreover, smeared cross section 
is used for finite energy resolution.
We show that ISR+FSR and smearing reduce cross sections
remarkably depending on the smearing parameter due to 
narrow decay width of bileptons. We obtain realistic 
discovery contours of couplings and masses for the lepton
flavour conserving and violating processes.
\end{abstract}


\maketitle

\section{Introduction}
In the minimal Standard Model (SM) the usual Higgs mechanism 
responsible for the electroweak symmetry breaking implies the 
conservation of the lepton number separately for each generation.
As is well known, the current low energy phenomenology of the SM 
is quite consistent with all present experiments. However, there 
has been no experimental evidence for the existence of the SM 
Higgs boson. This is one of the good reasons that other symmetry 
breaking mechanisms and extended Higgs sectors have not been 
excluded in the theoretical point of view. In addition, 
indication for neutrino oscillations necessarily violates 
lepton-flavour symmetry\cite{fukuda}.
In the theories beyond the SM doubly charged, lepton flavour 
changing, exotic bosons may occur.  Models with extended 
Higgs sectors include doubly charged Higgs boson \cite{haber}.
In the  supersymmetric extensions, such as SO(10) SUSY GUT model, 
supersymmetric lepton partners induce lepton 
flavor violation \cite{barbieri}.
The purpose of this paper is to study doubly charged bilepton 
search through resonance channel in $e^{-}e^{-}$ scattering
in some detail including electromagnetic initial+final state 
radiative corrections and smearing effects.
Bileptons  are defined as bosons carrying lepton 
number L=2 or 0 which  couple to two standard model 
leptons but not to quarks. Bileptons appear in
left-right symmetric models \cite{mohapatra} 
and also in models where 
$\text{SU(2)}_{\text{L}}$ gauge group is extended 
to SU(3) \cite{frampton}. Grand unified theories, 
technicolor and composite models predict 
the existence of bileptons as well as 
other exotic particles \cite{ross}. Classification and interactions of 
bileptons are provided by several works \cite{rizzo} and  
a comprehensive review has been presented  in \cite{cuypers}
including low and high energy bounds on their couplings.
Indirect constraints on the masses and couplings of doubly 
charged bileptons have been obtained from $\mu$ and $\tau$ 
physics,  muonium-antimuonium conversion and Bhabba scattering 
experiments \cite{swartz,sasaki,chang,willmann,okan,opal}. 
A search for doubly charged bilepton  
has been performed by DELPHI collaboration at LEP
\cite{delphi}

General effective lagrangian describing interactions of 
bileptons with the standard model leptons is generated by 
requiring $\text{SU(2)}_{\text{L}}\times \text{U(1)}_
{\text{Y}}$ invariance. We consider  the lagrangian involving 
the bilepton couplings to leptons only for L=2 bileptons as follows: 

\begin{eqnarray}
{\cal L}_{L=2}=&&g_1^{ij} \bar{\ell}_{i}^{c}i\sigma_2 
\ell_{j} L_1+\tilde{g}_1^{ij}
\bar{e}_{iR}^{c}e_{jR}\tilde{L}_1 \nonumber \\
+&&g_2^{ij}\bar{\ell}_{i}^{c}i\sigma_2\gamma_{\mu}
e_{jR}L_{2}^{\mu}+g_3^{ij}\bar{\ell}_{i}^{c}i\sigma_2\vec{\sigma}
\ell_{j}~.\vec{L}_3+h.c.
\end{eqnarray}

In the notations we have used ${\ell}$ is the left handed 
$\text{SU(2)}_\text{L}$ lepton doublet and $e_{R}$ is the 
right handed charged singlet lepton. Charge conjugate fields are 
defined as $\psi^{c}=C\bar{\psi}^T$ and $\sigma_1$, 
$\sigma_2$, $\sigma_3$ are the Pauli matrices.  
The subscript of bilepton fields $L_{1,2,3}$ and 
couplings $g_{1,2,3}$ denote $\text{SU(2)}_\text{L}$ singlets, 
doublets and triplets. We denote flavour indices by $i,j=1,2,3$

Here we are interested only in doubly charged bileptons. 
In order to express the lagrangian in terms of individual electron, 
bileptons and helicity projection operators 
$\text{P}_{\text{R/L}}=\frac{1}{2}(1\pm\gamma_5)$
we expand the Pauli matrices and lepton doublets
and write the lagrangian as :
\begin{eqnarray}
{\cal L}_{L=2}=&&\tilde{g}_1~\tilde{L}_1^{++}~\bar{e}^{c}P_R e
+g_2~ L_{2\mu}^{++}~\bar{e}^{c}\gamma^{\mu}P_L e
\nonumber\\
&&-\sqrt{2}g_3~ L_3^{++}~\bar{e}^{c}P_L e + h.c.
\end{eqnarray}
where superscripts of bileptons stand for their electric charges
and flavour indices have been skipped. 
When the scalar $L_{3}$ gains a vacuum expectation value it 
becomes a doubly charged Higgs that apears in left-right 
symmetric models.

$e^{-}e^{-}$ colliders\cite{ee}  are to be considered as a component 
of future linear $e^{+}e^{-}$ collider programs \cite{lc}. The initial 
state quantum numbers of $e^{-}e^{-}$ colliders make them 
suitable to probe resonances for doubly charged bileptons.

\section{Cross Sections for $e^{-}e^{-}\to e^{-}e^{-}$ 
and $e^{-}e^{-}\to \mu^{-}\mu^{-}$ resonant states }

 The unpolarized differential cross section for the process
 $e^{-}e^{-}\to e^{-}e^{-}$ including doubly charged scalar bilepton
  $\text{L}_{3}^{--}$ exchange with lepton flavour conserving 
couplings is given by
 
\begin{eqnarray}
 \frac{d\sigma}{d\cos{\theta}}=&&\frac{1}{2}(\frac{1}{32\pi s})
 \frac{1}{4}[|M(LL;LL)|^{2}+|M(RR;RR)|^{2}+|M(LR;LR)|^{2}
 \nonumber \\
+&&|M(RL;RL)|^{2}+|M(LR;RL)|^{2}+|M(RL;LR)|^{2}]
\end{eqnarray}

where the helicity amplitudes are given 
in terms of mandelstam invariants s, t and u as below
 
\begin{eqnarray}
M(LL;LL)&&=-2 g_{e}^{2}s (\frac{1}{t}+\frac{1}{u})-
2 C_{L}^{2}s (\frac{1}{t-M_{Z}^{2}}+\frac{1}{u-M_{Z}^{2}})
-g_{L}^{2}\frac{s}{s-M_{L}^{2}+iM_{L}\Gamma_{L}} \\
M(RR;RR)&&=-2 g_{e}^{2}s (\frac{1}{t}+\frac{1}{u})-
2 C_{R}^{2}s (\frac{1}{t-M_{Z}^{2}}+\frac{1}{u-M_{Z}^{2}}) \\
M(LR;LR)&&=-2 g_{e}^{2} \frac{u}{t}+2C_{L}C_{R}\frac{u}{t-M_{Z}^{2}}
\\
M(RL;RL)&&=M(LR;LR)\\
M(LR;RL)&&=-2 g_{e}^{2} \frac{t}{u}-2C_{L}C_{R}\frac{t}{u-M_{Z}^{2}}
\\
M(RL;LR)&&=M(LR;RL)
\end{eqnarray}

In the above expressions we denote bilepton-lepton-lepton 
coupling by $g_{L}$ where subscript L indicates the bilepton. 
$M_{L}$ and  $\Gamma_{L}=g_{L}^{2}M_{L}/(16\pi)$ are scalar 
bilepton mass and decay width into leptons, respectively.
In this work, we consider that doubly charged bileptons 
can decay only to leptons.
The mandelstam invariant s is defined as 
the square of total energy of incoming particles in the center 
of mass system (c.m.) and other variables t and u can be written
in terms of angle  between incoming and outgoing leptons in c.m. system 

\begin{eqnarray}
t=-\frac{s}{2}(1-\cos{\theta})~,~~~
u=-\frac{s}{2}(1+\cos{\theta})
\end{eqnarray}

The couplings $C_{L}$ and $C_{R}$ can be connected to  the 
electromagnetic coupling parameter 
$g_{e}^{2}=4\pi\alpha_{em}$ and Weinberg angle 
$\theta_{W}$ 

\begin{eqnarray}
C_{L}&&=\frac{g_{z}}{2}(C_{V}+C_{A})~,~~~
C_{R}=\frac{g_{z}}{2}(C_{V}-C_{A})\\
C_{V}&&=2\sin^{2}{\theta_{W}}-\frac{1}{2}~,~~~C_{A}=-\frac{1}{2}
~~~\text{for}~~e,~\mu~,\tau \\
g_{z}&&=\frac{g_{e}}{\sin{\theta_{W}}\cos{\theta_{W}}}
\end{eqnarray}

In the case of the process $e^{-}e^{-}\to \mu^{-}\mu^{-}$  
only flavour violating  doubly charged  bilepton couplings 
contribute to the cross section via s channel resonant 
diagram. Therefore, it is enough to remove t and u channel 
contribution from the above cross section. 
For this case the decay width $\Gamma_{L}$ into leptons 
must be enlarged.

\section{Initial and Final State Electromagnetic 
Radiative Corrections}

Due to small mass of the electron, a significant role
is played by the electromagnetic radiative corrections
to the initial electron-positron state epecially at
linear collider  energies.
In this work we use structure function formalism to describe
the electromagnetic radiative corrections in $e^{+}e^{-}$
colliders \cite{fadin} .
In the case of $e^{-}e^{-}\to \mu^{-}\mu^{-}$ process 
the cross section can be written in the following form within
this formalism

\begin{eqnarray}
\sigma(s)=\int{dx_{1}}\int{dx_{2}}\int{dx_{3}}\int{dx_{4}}
~ D_{1}(x_{1},s)~D_{2}(x_{2},s) ~
D_{3}(x_{3},s^{\prime\prime})~D_{4}(x_{4},s^{\prime\prime}) 
\sigma^{\prime}(s^{\prime})
\end{eqnarray}
where $\sigma^{\prime}(s^{\prime})$ is the cross section
with reduced energy $s^{\prime}=x_{1}x_{2}s$.
$D_{1}(x_{1},s)$ and  $D_{2}(x_{2},s)$ stand for the initial electron
structure function giving the probability of finding
an electron within an electron  with
longitudinal momentum fractions $x_{1}$ and $x_{2}$.
$D_{3}(x_{3},s^{\prime\prime})$ and $D_{4}(x_{4},s^{\prime\prime})$
represent structure functions for the final leptons  with
longitudinal momentum fractions $x_{3}$ and $x_{4}$. Here 
$s^{\prime\prime}$ is defined as $s^{\prime\prime}=x_{3}x_{4}s^{\prime}$.
Although several definitions of the structure functions are present
we use the following ones which are used by HERWIG(a multipurpose 
Monte Carlo event generator which has been extensively used at CERN 
LEP) \cite{herwig}

\begin{eqnarray}
D(x,Q^{2})&&=\beta (1-x)^{\beta-1} g(x,Q^{2})
\\
g(x,Q^{2})&&=e^{\beta (1+x/2)x/2} (1-\beta^{2} \frac{\pi^{2}}
{12})+y\frac{\beta^{2}}{8} y [(1+x)\{(1+x)^{2}+3 \log{x}\}
-\frac{4 \log{x}}{1-x}]
\\
y&&=[\beta (1-x)^{\beta-1}]^{-1} ~,~~~
\beta=\frac{\alpha_{em}}{\pi}(\log{\frac{Q^{2}}{M^{2}}-1})
\end{eqnarray}

In expression $\beta$ the value of $Q^{2}$ and M take the 
$s(s^{\prime\prime})$ and initial(final) lepton mass depending on 
initial(final) state structure function.

To avoid divergency at the upper limit of the momentum fraction,
$x=1$, the cross section can be transformed into different form

\begin{eqnarray}
\sigma(s)&&=\int_{0}^{1-\epsilon}{dx_{1}}{dx_{2}}{dx_{3}}{dx_{4}}
D_{1}(x_{1},s) D_{2}(x_{2},s) D_{3}(x_{3},s^{\prime\prime}) 
D_{4}(x_{4},s^{\prime\prime})
\sigma^{\prime}(s^{\prime}) \nonumber \\
&&+\frac{1}{6}\sum_{i,j,k,\ell}\int_{0}^{1-\epsilon}
{dx_{i}}{dx_{j}}{dx_{k}}~D_{i}(x_{i}) D_{j}(x_{j}) D_{k}(x_{k})
\epsilon^{\beta_{\ell}}
g_{\ell}(x_{\ell})|_{x_{\ell}=1} \nonumber \\
&&+\frac{1}{4}\sum_{i,j,k,\ell}\int_{0}^{1-\epsilon}
{dx_{i}}{dx_{j}}~D_{i}(x_{i}) D_{j}(x_{j}) 
\epsilon^{\beta_{k}}~ \epsilon^{\beta_{\ell}}
g_{k}(x_{k})~g_{\ell}(x_{\ell})|_{{x_{k}=1},{x_{\ell}=1}}\nonumber \\
&&+\frac{1}{6}\sum_{i,j,k,\ell}\int_{0}^{1-\epsilon}
{dx_{i}}~D_{i}(x_{i})
\epsilon^{\beta_{j}}~\epsilon^{\beta_{k}}~ \epsilon^{\beta_{\ell}}
g_{j}(x_{j})~g_{k}(x_{k})~g_{\ell}(x_{\ell})
|_{{x_{j}=1},{x_{k}=1},{x_{\ell}=1}}\nonumber \\
&&+\epsilon^{2\beta_{1}}~\epsilon^{2\beta_{3}}
g_{1}(x_{1},s)~g_{2}(x_{2},s)~g_{3}(x_{3},s^{\prime\prime})~
g_{4}(x_{4},s^{\prime\prime})
|_{{x_{1}=1},{x_{2}=1},{x_{3}=1},{x_{4}=1}}
\end{eqnarray}

where $\epsilon$ can be taken as $10^{-9}-10^{-12}$. In this
region of $\epsilon$ the cross section changes by a factor of 0.99.
If one takes smaller $\epsilon$ values,  higher machine precision
gives softer $\epsilon$ dependence. The subscript of 
$\beta$ inside integrands is used to clarify the difference between initial 
and final state $\beta$ values where $\beta_{1}=\beta_{2}$ stand 
for initial and $\beta_{3}=\beta_{4}$ for final leptons. 
The following transformation gives relatively smooth integrand

\begin{eqnarray}
&&\int_{x_{10}}^{1-\epsilon}{dx_{1}}\int_{x_{20}}^{1-\epsilon}
{dx_{2}} \int_{x_{30}}^{1-\epsilon}{dx_{3}}
\int_{x_{40}}^{1-\epsilon}{dx_{4}}
D_{1}(x_{1},s) D_{2}(x_{2},s) D_{3}(x_{3},s^{\prime\prime})
D_{4}(x_{4},s^{\prime\prime})\sigma^{\prime}(s^{\prime})=
\nonumber \\
&&\int_{E_{1min}}^{E_{1max}}{dE_{1}} \int_{F_{1min}}^{F_{1max}}
{dF_{1}}\int_{E_{2min}}^{E_{2max}}{dE_{2}}\int_{F_{2min}}^{F_{2max}}
{dF_{2}}~g_{1}(x_{1},s)
~g_{2}(x_{2},s)~g_{3}(x_{3},s^{\prime\prime})~
g_{4}(x_{4},s^{\prime\prime})
\sigma^{\prime}(s^{\prime})
\end{eqnarray}

where

\begin{eqnarray}
&&x_{10}=\frac{\tau_{min}}{(1-\epsilon)^{3}}~~,~~~
x_{20}=\frac{\tau_{min}}{x_{1}(1-\epsilon)^{2}}~~,~~~\nonumber \\
&&x_{30}=\frac{\tau_{min}}{x_{1}x_{2}(1-\epsilon)}~~,~~~
x_{30}=\frac{\tau_{min}}{x_{1}x_{2}x_{3}}
\end{eqnarray}

\begin{eqnarray}
&&x_{1}=1-(-E_{1})^{1/\beta_{1}}~~,~~~
E_{1min}=-(1-x_{10})^{\beta_{1}}~~,~~
E_{1max}=-\epsilon^{\beta_{1}} \nonumber \\
&&x_{2}=1-(-F_{1})^{1/\beta_{2}}~~,~~~
F_{1min}=-(1-x_{20})^{\beta_{2}}~~,~~
F_{1max}=-\epsilon^{\beta_{2}}\nonumber \\
&&x_{3}=1-(-E_{2})^{1/\beta_{3}}~~,~~~
E_{2min}=-(1-x_{30})^{\beta_{3}}~~,~~
E_{2max}=-\epsilon^{\beta_{3}} \nonumber \\
&&x_{4}=1-(-F_{2})^{1/\beta_{4}}~~,~~~
F_{2min}=-(1-x_{40})^{\beta_{4}}~~,~~
F_{2max}=-\epsilon^{\beta_{4}}
\end{eqnarray}

The parameter $\tau_{min}$ is a square of  minimum energy fraction 
carried by final state leptons. Although it depends on the 
experimental conditions it can be taken as 
$\tau_{min}=4M_{f}^{2}/s$ for theoretical purpose.  

In the case of $e^{-}e^{-}\to e^{-}e^{-}$ scattering the 
difficulty arises due to presence of t-channel Standard Model 
processes. This is the two scale problem. The difficulty 
can be handled by the following idea.
The s-channel is dominant only around the 
bilepton resonance region, whereas the one photon  t-channel 
exchange dominates the cross section away from  
resonance region. Moreover, at large scattering angles 
the scale t becomes of the same order as scale s.
Therefore, in the region of resonance at large angles,
$e^{-}e^{-}\to e^{-}e^{-}$ scattering can be considered as 
one scale problem and the above calculation for initial and 
final state QED radiative correction is applicable in both 
cases.

\section{Smeared Cross Section and Discussion}

In order to account for a finite resolution in the invariant 
mass of final state leptons $M_{\ell\ell}$, we consider a 
smeared cross section as defined below

\begin{eqnarray}
\sigma(s_{0})=\int_{E_{0}-D/2}^{E_{0}+D/2}
\frac{dE}{\sqrt{2\pi}D}\exp{[-\frac{(E_{0}-E)^{2}}{2D^{2}}]}
\sigma(s)
\end{eqnarray}
where $E_{0}=\sqrt{s_{0}}$ and $E=\sqrt{s}$.
Smearing parameter D corresponds to an overall energy resolution.
The effect of ISR, FSR and smearing can be seen from Table~\ref{tab1}
which shows the significance 
\begin{eqnarray}
S=\frac{|\sigma-\sigma_{SM}|}{\sqrt{\sigma_{SM}}}\sqrt{L_{int}}
\end{eqnarray}
for the process $e^{-}e^{-}\to e^{-}e^{-}$
at a  resonance point with an assumed bilepton mass  $M_{L}=500$
GeV and an integrated luminosity $L_{int}=10000pb^{-1}$.
In previous works,
indirect  upper limits on the bilepton-lepton-lepton 
couplings were found to be of the order of $O(10^{-1})$ using 
CERN LEP data for the bilepton masses 200-800 GeV \cite{okan,opal}.
For comparison we take three values of coupling 
 $g_{L}=0.1, 0.01, 0.001$ and two values of the 
smearing parameter D=0.05, 5 GeV. Depending on the couplings and 
bilepton mass, the values of the bilepton decay width to leptons 
are also exhibited in the table to compare with D parameter.
It is clear from the values of the significance $S_{0}$ 
that the cross section is independent of the coupling at 
the resonance point when ISR, FSR and smearing are not included.
Since we consider only lepton decay channels of bileptons
the decay width is very narrow for small couplings and for the
masses specified above. Therefore, there is a difficulty 
to observe invariant masses of two leptons $M_{\ell\ell}$ 
experimentally. If the experimental energy resolution is 
around D=0.01E (D=5 GeV) the decay width is 
$\Gamma_{L}<<D$ and the significance drastically reduces due 
to ISR, FSR and smearing. Moreover, for the coupling 
$g_{L}=0.001$ the peak is unobservable because of $S_{2}<<1$.
If we assume higher energy resolution such as
D=0.0001E (D=0.05 GeV) the bilepton decay width 
$\Gamma_{L}>D$ for only $g_{L}=0.1$ and there is no need
to consider smeared cross section. However, for smaller 
couplings $S_{2}$ needs to be considered. In this case,
the bileptons can be observed also for the coupling 
$g_{L}=0.001$.

\begin{figure}
\includegraphics{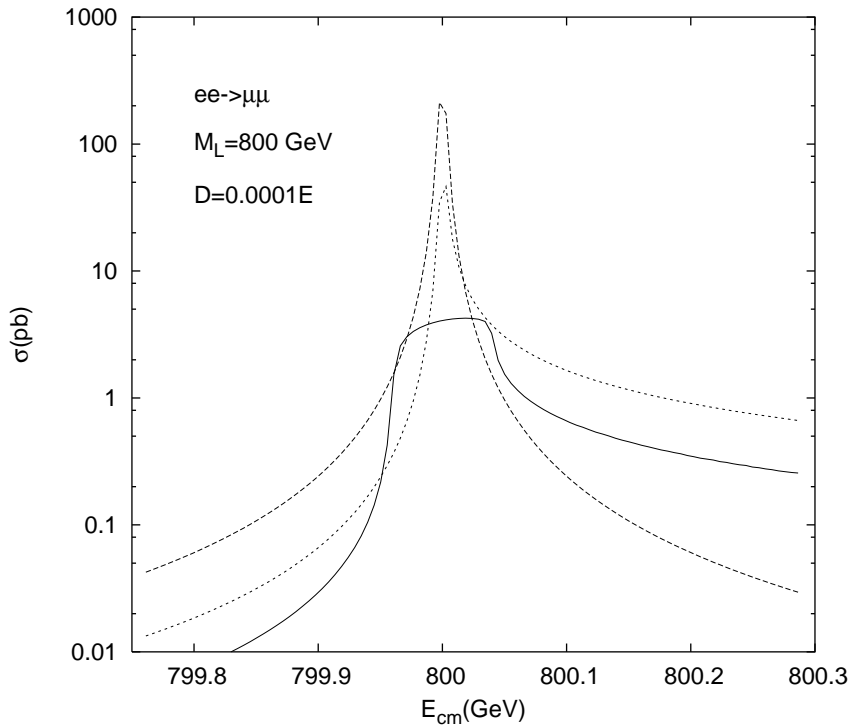}
\caption{Bilepton resonance curves from the process
$e^{-}e^{-} \to \mu^{-}\mu^{-}$ assuming bilepton mass is
$M_{L}=800$ GeV. The highest curve around the resonance point
is free of initial, final state radiative corrections(ISR, FSR) and
smearing effect on the cross section. The second curve includes
ISR and FSR. The lowest curve covers ISR, FSR and smearing effect
with smearing parameter D=0.0001E.
\label{fig1}}
\end{figure}

\begin{figure}
\includegraphics{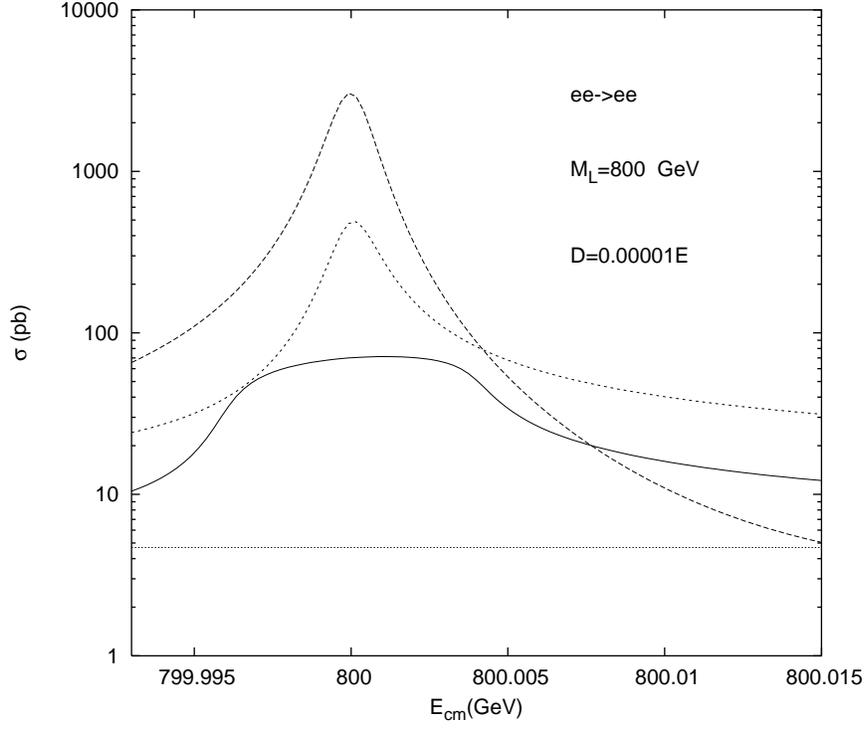}
\caption{Bilepton resonance curves from the process
$e^{-}e^{-} \to e^{-}e^{-}$ assuming bilepton mass is
$M_{L}=800$ GeV. The highest curve around the resonance point
is free of initial, final state radiative corrections(ISR, FSR) and
smearing effect on the cross section. The second curve includes
ISR and FSR. The third curve covers ISR, FSR and smearing effect
with smearing parameter D=0.00001E. The lowest curve is the
Standard Model background.
\label{fig2}}
\end{figure}

Similar features are shown  in Fig.~\ref{fig1}
for the resonance in $e^{-}e^{-}\to \mu^{-}\mu^{-}$ process 
at a resonance point $E_{cm}=M_{L}=800$ GeV
with the smearing parameter $D=0.0001E$. Here we assume
lepton flavour number is violated.
The bilepton-lepton-lepton coupling $g_{L}=0.01$ and  
$\tau_{min}=0.01$  are taken into account.  
The cross section at the peak of the resonance is about 
200pb without  ISR, FSR and smearing. ISR and FSR reduce 
the cross section to 46pb  whereas the smearing 
in addition to ISR+FSR reduces it  to 4pb,
i.e. almost 50 times smaller than the 
highest point. 
If we consider lepton flavour  is conserved the 
resonance peak should be observed in the process 
$e^{-}e^{-}\to e^{-}e^{-}$  with the SM background of 
Moeller scattering. In addition to $g_{L}=0.01$ and
$\tau_{min}=0.01$ we apply a cut on the scattering angle 
$|\cos{\theta}|=0.8$ to avoid t-channel instabilities. 
In this case, bilepton decay width to leptons 
is highly narrower than the case of lepton flavour violation.
More clearly, Fig.~\ref{fig2} shows this behaviour of 
cross sections at an assumed 
resonance point ($M_{L}=800$ GeV) with and without ISR+FSR and 
smearing including Standard Model background (Moeller scattering). 
With a smearing parameter $D>0.0001E$ the peak in the 
invariant mass of $M_{ee}$ is hardly observable after smearing 
due to SM background. For illustration 
we use D=0.00001E in Fig.~\ref{fig2}.
Therefore, we will need  high resolution 
detectors to discover bileptons from this process.

\begin{figure}
\includegraphics{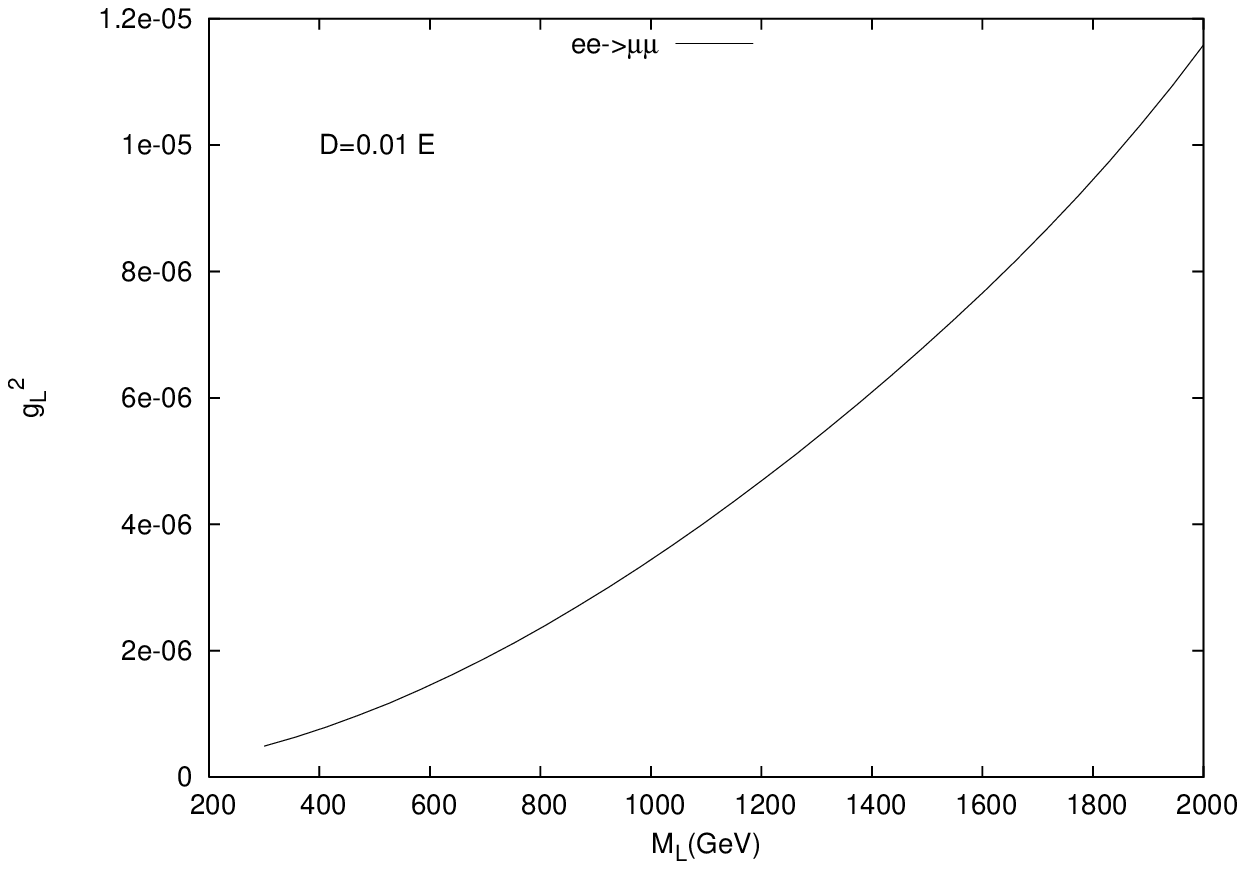}
\caption{$\%95$ C.L. contour of the bilepton-lepton-lepton
coupling squared $g_{L}^{2}$ and bilepton mass $M_{L}$ coming from
lepton flavour violating $e^{-}e^{-} \to \mu^{-}\mu^{-}$ scattering.
Each curve has been  obtained by considering ISR, FSR and
smeared cross section with smearing parameter D=0.01E.
The left part of the each curve is the allowed region.
\label{fig3}}
\end{figure}

\begin{figure}
\includegraphics{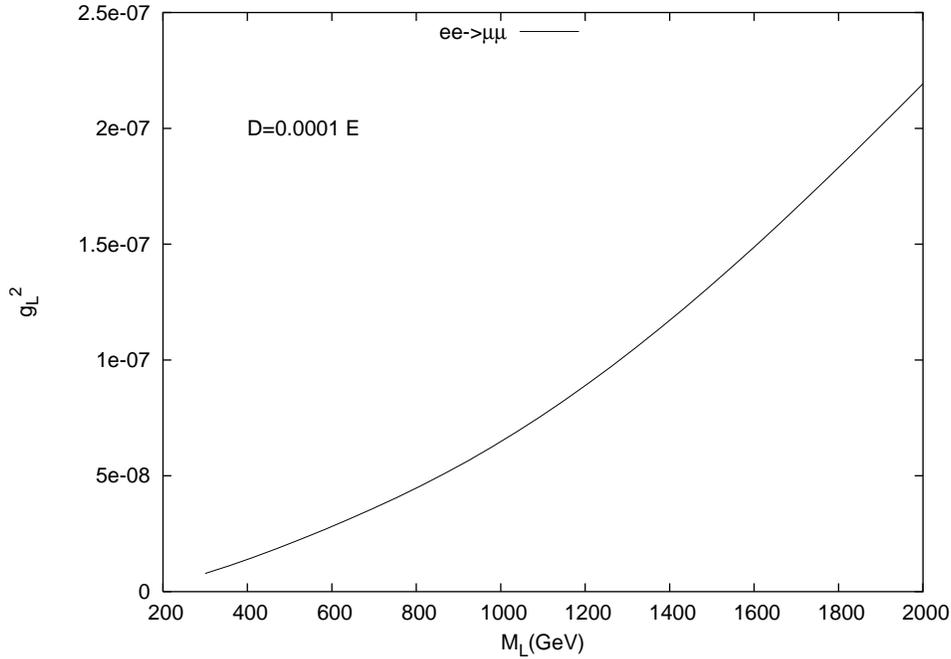}
\caption{The same as the previous figure but for
the smearing parameter D=0.0001E.
\label{fig4}}
\end{figure}

\begin{figure}
\includegraphics{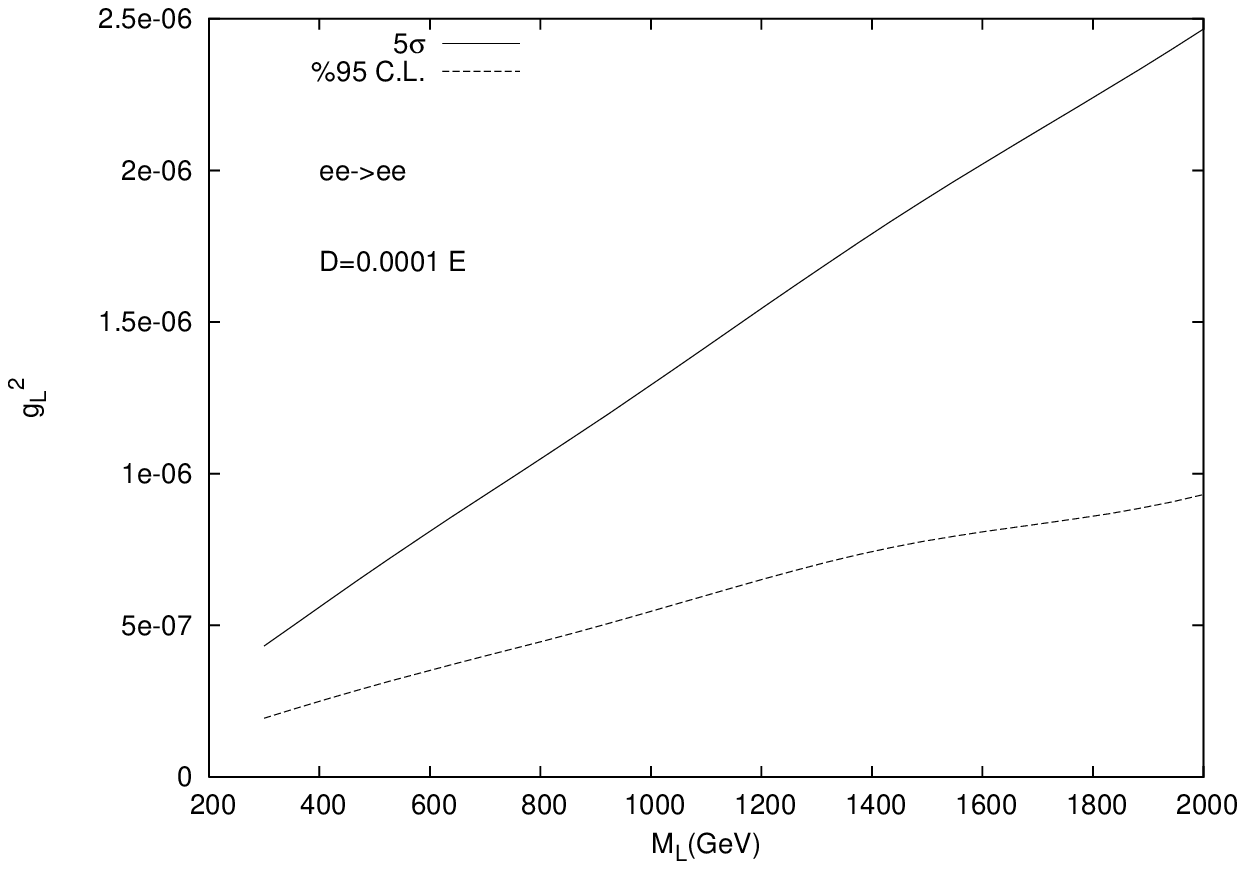}
\caption{$5\sigma$ and $\%95$ C.L. contour of the bilepton-lepton-lepton
coupling squared $g_{L}^{2}$ and bilepton mass $M_{L}$ coming from
lepton flavour conserving $e^{-}e^{-} \to e^{-}e^{-}$ scattering.
Each curve has been  obtained by considering ISR, FSR and
smeared cross section with smearing parameter D=0.0001E.
The left part of the each curve is the allowed region.
\label{fig5}}
\end{figure}

In order to estimate discovery contour of  $g_{L}^{2}$ and 
$M_{L}$ we use one parameter one sided $\chi^{2}$ analysis 
at $\%95$ C.L. and $5\sigma$ significance for 
$e^{-}e^{-}\to e^{-}e^{-}$ process. 
In the case of lepton flavour violating process 
$e^{-}e^{-}\to \mu^{-}\mu^{-}$ there is no SM background 
and all the events consist of signal events. Then,
$\%95$ C.L. contour can be obtained by taking 
Poisson variable as the observed 
events with Poisson mean $\nu=9.15$.
In Fig.~\ref{fig3} and Fig.~\ref{fig4} discovery contour of 
$g_{L}^{2}$ and $M_{L}$ are plotted from the 
lepton flavour violating process including ISR+FSR and smearing  
with smearing parameter D=0.01E  and D=0.0001E.
Fig.~\ref{fig5} shows the similar curves for the lepton 
flavour conserving process with D=0.0001E.
In all these figures the left part of the each curve 
is the allowed region. Certainly, higher luminosity 
creates more allowed region.

\begin{table}
\caption{ Comparison of the significances  
$S_{0}$, $S_{1}$ and $S_{2}$ for the process 
$e^{-}e^{-}\to e^{-}e^{-}$ 
at an assumed resonance point  $M_{L}=500$ GeV to demonstrate 
the influence of the initial, final state electromagnetic radiative 
corrections and smearing effect. $S_{1}$ is  the significance 
with ISR+FSR and  $S_{2}$ with ISR+FSR and smearing. 
$S_{0}$ does not cover any corrections and smearing.
\label{tab1}}
\begin{ruledtabular}
\begin{tabular}{cccccc}
$g_{L}$ & D(GeV) &$\Gamma_{L}$(GeV) & $S_{1}$& $S_{2}$ &$S_{0}$  \\
\hline
0.1  &   5    & 0.1  & 40000  & 1400   & 200900\\
0.01 &   5    & 0.001  & 19800  & 10  & 200900\\
 0.001 & 5   & 0.00001  & 9700  & 0.04   & 200900\\
0.1 & 0.05   & 0.1  &  40000  & -  & 200900\\
0.01& 0.05 & 0.001 &  19800 & 700& 200900\\
0.001 & 0.05 &  0.00001 & 9700   & 5  & 200900\\
\end{tabular}
\end{ruledtabular}
\end{table}

\end{document}